\documentclass[twocolumn,showpacs,pra,a4paper]{revtex4}

\usepackage{amssymb}
\usepackage{graphicx}
\usepackage{color}
\newcommand{\beq}{\begin{equation}}
\newcommand{\eeq}{\end{equation}} \newcommand{\bqa}{\begin{eqnarray}} \newcommand{\eqa}{\end{eqnarray}} \newcommand{\beqa}{\begin{eqnarray}} \newcommand{\eeqa}{\end{eqnarray}} \newcommand{\beqan}{\begin{eqnarray*}} \newcommand{\eeqan}{\end{eqnarray*}}   \newcommand{\eq}[1]{Eq.\ (\ref{#1})}  \newcommand{\bra}[1]{\langle{#1}|} \newcommand{\ket}[1]{|{#1}\rangle}   \newcommand{\ip}[1]{\langle{#1}\rangle}     \newcommand{\half}{\frac{1}{2}} \newcommand{\sbin}[2]{\left({}^{#1}_{#2}\right)}

\begin{document}

\title{Beyond Landauer erasure}

\author{Stephen M. Barnett$^{1}$ and Joan A. Vaccaro$^{2}$}

\affiliation{%
 $^{1}$ School of Physics and Astronomy, University of Glasgow, Glasgow G12 8QQ, United Kingdom\\
 $^{2}$ Centre for Quantum Computation and Communication Technology (Australian Research Council), Centre for Quantum Dynamics, Griffith University, Brisbane, Queensland 4111 Australia }


\begin{abstract}
In thermodynamics one considers thermal systems and the maximization of entropy subject to the conservation of energy.  A consequence is Landauer's erasure principle, which states that the erasure of 1 bit of information requires a minimum energy cost equal to $kT\ln(2)$ where $T$ is the temperature of a thermal reservoir used in the process and $k$ is Boltzmann's constant.  Jaynes, however, argued that the maximum entropy principle could be applied to any number of conserved quantities which would suggest that information erasure may have alternative costs. Indeed we showed recently that by using a reservoir comprising energy degenerate spins and subject to conservation of angular momentum, the cost of information erasure is in terms of angular momentum rather than energy.  Here we extend this analysis and derive the minimum cost of information erasure for systems where different conservation laws operate.  We find that, for each conserved quantity, the minimum resource needed to erase 1 bit of memory is $\lambda^{-1}\ln(2)$ where $\lambda$ is related to the average value of the conserved quantity. The costs of erasure depend, fundamentally, on both the nature of the physical memory element and the reservoir with which it is coupled.  
\end{abstract}



\maketitle

\section{Introduction} \label{sec:intro}

The idea of a link between information and thermodynamics can be traced back to Maxwell's demon, a supposed microscopic intelligent being, the actions of which might present a challenge to the second law of thermodynamics \cite{Maxwell,Leff}.  This idea was made quantitative by Szilard who showed, by means of a simple one-molecule gas, that information acquisition, for example by a Maxwell demon, is necessarily accompanied by an entropy increase of not less than $k\ln(2)$ \cite{Szilard}, where $k$ is Boltzmann's constant. A closely related phenomenon is the demonstration, due to Landauer, that erasing an unknown bit of information requires energy to be dissipated as heat, amounting to not less than $kT\ln(2)$ \cite{Landauer,PleVit,Maroney}, where $T$ is the temperature of the environment surrounding the bit.

Information theory is usually cast in a form that is independent of any particular physical realization.  In particular, information processing could take place within degenerate manifolds of identical energy. In such a situation, it is difficult to see what role the energy could play. It would then be natural to question the place of Landauer's erasure principle relying, as it does, on the thermal \emph{energy} of the surrounding reservoir. Indeed we have shown that information can be erased at a cost of spin angular momentum and not energy \cite{PRSA}. We explore this idea further here. We show, in particular, that the energy does not have a special place among conserved quantities. Rather, Landauer's principle needs to be supplemented with costs associated with each of the conserved quantities.

The theoretical foundation for our new results lies in Jaynes' maximum entropy principle \cite{Jaynes,Jaynes Book}, which is colloquially called the {\it MaxEnt} principle. It applies when we have only partial information about a system.  The assignment of a probability distribution to account for the missing information that maximizes the entropy ``...is the least biased estimate possible on the given information; i.e. it is maximally noncommittal with respect to the missing information'' \cite{Jaynes}. It has repercussions for erasure, because the entropy of the environment increases as it absorbs the entropy of the erased information, and any physical resources linked to the entropy change as a result.

The organisation of this paper is as follows.  In section 2 we review the MaxEnt principle in as much as it applies to information erasure.  We then review an archetypical erasure model in section 3. The main results, the minimum cost of erasure in terms of a number of conserved quantities, are presented in section 4 and we end with a discussion in section 5.

\section{General principles for information erasure}

To perform information erasure we need a system, which we shall call a reservoir, that is capable of absorbing the entropy of the unwanted information.  Typically this means that the reservoir will be large in that it will have very many degrees of freedom. The kind of reservoir we wish to analyze differs from the familiar thermal reservoir in that it is not necessarily characterised by an average energy, or temperature.  Instead, the only information we have about the reservoir is the expectation values for a number of physical variables.  Let these expectation values be written as $\ip{\hat V_k}$ where $\hat V_k$ is a physical variable, such as energy or angular momentum etc., and $k$ is an integer which indexes different variables. We assume that the variables are independent and so in quantum theory they are represented by commuting self-adjoint operators. Then according to the MaxEnt principle, the best description of the state of the system is one for which entropy is maximized, in which case the density operator for the reservoir is given by
\beq
   \hat \rho = \exp\left[-\left(\lambda_0+\sum_k\lambda_k\hat V_k\right)\right]
\eeq
and its Shannon entropy is \cite{Jaynes}
\beq
    S_r = \lambda_0+\sum_k\lambda_k\ip{\hat V_k}\ .
\eeq
Here $\lambda_k$ are Lagrange multipliers.  If the system is altered in such a way that the expectation values $\ip{\hat V_k}$ and operators $\hat V_k$ change independently of each other then a small change in entropy is given by \cite{Jaynes}
\beq
    \delta S_r = \sum_k\lambda_k(\delta\ip{\hat V_k}-\ip{\delta\hat V_k})\ .
\eeq
Here $\delta\ip{\hat V_k}$ denotes a change in the expectation value of $\hat V_k$ whereas $\ip{\delta\hat V_k}$ is the expectation value of a change in the operator $\hat V_k$ itself.
The variables are not necessarily conserved, but rather their expectation values are presumed to be known at all times.

Let the memory be a two-state system which is initially in a maximally mixed state with a Shannon entropy of $S_{m}=\ln 2$.  Erasing the information represented by this state will entail a physical process that leaves the memory is a predetermined pure state and thus the memory will suffer a change in entropy of $\delta S_{m}=-\ln 2$.  If the erasure process is reversible, the total entropy of the combined system will be unchanged and so
\beq
    0= \delta S_r+\delta S_m\ .
\eeq
Given $\delta S_{m}=-\ln 2$, the entropy of the reservoir will necessarily be increased by $\delta S_{r}=\ln 2$ in the process. If the erasure leaves the variables $\hat V_k$  unchanged (i.e. $\delta \hat V_k=0$), then the only way an entropy increase can be accommodated is by an increase in the expectation values as follows
\beq
    \delta S_r = \sum_k \lambda_k\delta\ip{\hat V_k}
\eeq
which becomes
\beq
    \ln 2 = \sum_k \lambda_k\delta\ip{\hat V_k}\ .
    \label{eq:general cost}
\eeq
This result shows that the reservoir undergoes physical changes as a result of the erasure. In particular, the changes to the variables $\ip{\hat V_k}$ represent physical costs of the process. There is no distinction between the relative sizes of the cost associated with each variable; the only requirement is that the weighted sum of the costs be equal to $\ln 2$.  The cost of erasure could be paid in terms of just a single variable, say $k=1$, in which case
\beq
    \delta\ip{\hat V_1}=\frac{1}{\lambda_1}\ln 2\ .
\eeq
If $\hat V_1$ represents the Hamiltonian $\hat H$ then $\lambda_1=1/kT$ where $k$ is Boltzmann's constant and $T$ is the temperature, and we recover Landauer's principle:
\beq
    \delta\ip{\hat H}=kT\ln 2\ .
\eeq
This need not be the case, however; the erasure mechanism we reported in Ref. \cite{PRSA} corresponds to $\hat V_1$ representing the $z$ component of spin angular momentum, $\hat J_z$, in which case the corresponding cost of erasure is in terms of spin angular momentum and not energy. More will be said about this result later.

\section{Archetypal erasure model} \label{sec:archetypal}

\begin{figure}  
\begin{center}
\includegraphics[width=60mm]{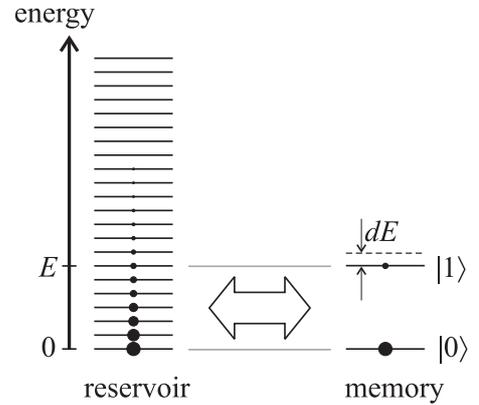}
\end{center}
\caption{Energy level diagram for the archetypal erasure model.  For clarity, only a few representative energy levels in the continuous energy spectrum of the reservoir are shown.  The reservoir and memory remain in thermal equilibrium and exchange energy while the energy gap between the states of the memory is slowly increased. The size of the circles on each level indicate the relative probabilities of the states. \label{fig1:energy}
}

\end{figure}

As the forgoing demonstrates, erasure entails transferring entropy from the memory to the reservoir.  A generic way to do this is as follows \cite{PRSA}.  Let the memory be represented by a two-state system with energy eigenstates $\ket{0}$ and $\ket{1}$, which we will refer to as logic states.  Next, imagine the memory is in an initially unknown state and that we wish to reset it by forcing it into the logic state $|0\rangle$. Let the two states $|0\rangle$ and $|1\rangle$ be initially degenerate, with energy $0$.  We can erase the memory by placing it in contact with a conventional thermal reservoir at temperature $T$ and then inducing an energy splitting between the states so that $|0\rangle$ has energy $0$, but the state $|1\rangle$ has energy $E$, as illustrated in Fig.~\ref{fig1:energy}.  The splitting is induced quasi-statically, that is, sufficiently slowly that the memory system is in thermal equilibrium with the thermal reservoir. The state of the memory when the energy splitting is $E$ is governed by the Boltzmann (or maximum entropy) distribution with density operator
\beq
   \hat\rho = \frac{\ket{0}\bra{0}+e^{-E/kT}\ket{1}\bra{1}}{1+e^{-E/kT}}\    .
\eeq
The work required to increase the splitting from $E$ to $E+dE$ while in contact with the reservoir is given the probability of occupation of the state $\ket{1}$ multiplied by $dE$, that is $dW=e^{-E/kT}(1+e^{-E/kT})^{-1}dE$. The total work in increasing the splitting from zero to infinity is $W=\int dW=kT\ln2$ and the final state of the memory system is the logic state $\ket{0}$ as required. The memory is removed from the reservoir and the energy degeneracy then restored. The erasure here is driven by maximizing the entropy subject to conservation of energy as the energy gap between the states of the memory system grows.

\section{Erasure with multiple costs} \label{sec:multiple}

We now use the principle underlying the archetypal erasure model to construct a more general erasure process.  Consider the situation in which the memory logic states $\ket{0}$ and $\ket{1}$ are associated with different eigenvalues of another conserved observable in addition to energy.  For definiteness, we take this observable to be the $z$ component of angular momentum and the memory to be a spin-$\half$ particle with $\ket{0}$ representing the eigenstate with eigenvalue $-\frac{1}{2}\hbar$ and $\ket{1}$ representing the eigenstate with eigenvalue $\frac{1}{2}\hbar$. Let the reservoir of the previous section be constructed of a collection of $N$ such particles.  Imagine that the information known about the reservoir are the expectation values of its total energy and the $z$ component of angular momentum, $\ip{\hat H}$ and $\ip{\hat J_z}$, respectively.  According to the MaxEnt principle, the best description of the state of the reservoir is given by
\beq
   \hat \rho = \exp[-(\mu+\beta \hat H +\gamma \hat J_z)]
   \label{eq:rho general spin}
\eeq
where $\mu$, $\beta$ and $\gamma$ are appropriate Lagrange multipliers.  We will investigate the situation where $\beta$ is independent of $\gamma$ in the following section.  For the moment, however, let us consider the situation where the internal spin and energy are correlated.  A case in point is where each spin particle experiences the same magnetic field orientated in the $z$ direction.  Then the logical states will be Zeeman shifted in energy with, for example, $\ket{1}$ being at an energy $\varepsilon$ higher than $\ket{0}$. In this case we can write the Hamiltonian $\hat H$ as a sum of an internal term $\hat H_{in}$ corresponding to the Zeeman shift and an external term $\hat H_{ex}$ corresponding to the motional degrees of freedom of the particles. Because $\hat H_{in}$ and $\hat J_z$ share the same eigenstates, we can write
\beq
   \hat H_{in}=\frac{\varepsilon}{\hbar}\hat J_z\ ,
   \label{eq:H = J}
\eeq
and it immediately follows that knowing both of $\ip{\hat H_{in}}$ and $\ip{\hat J_z}$ does not give any more information about the state of the reservoir than knowing just one.  In fact, in this case \eq{eq:rho general spin} reduces to
\beqa
   \hat \rho &=& \exp[-(\mu+\beta \hat H_{ex}+\gamma\hbar\hat H_{in }/\varepsilon)]\nonumber\\
   &=&\exp[-(\mu+\beta \hat H_{ex}+\gamma\hat J_z)]\ .
   \label{eq:rho correlated spin}
\eeqa
If we allow the internal spin states and external motion to be at the same temperature $T$, then $\beta=\gamma\hbar/\varepsilon=1/kT$ and
\beqa
   \hat \rho &=& \exp\{-[\mu+\beta (\hat H_{ex}+\hat H_{in})]\}\nonumber\\
   &=&\exp\{-[\mu+\beta (\hat H_{ex}+\varepsilon \hat J_z/\hbar)]\}\ .
   \label{eq:rho correlated spin- same temp}
\eeqa

It is convenient to use a set of basis states $\ket{n,\nu}$ to describe the internal collective spin state the reservoir in terms of the number of spins, $n$, in the logical state $\ket{1}$ (i.e. spin up) and the multiplicity index $\nu$ which specifies uniquely an arrangement of $n$ spins in the state $\ket{1}$ and $N-n$ in the state $\ket{0}$. The full basis set is given by $\{\ket{n,\nu_n}: n=0,\ldots,N; \nu_n=1,\ldots,\sbin{N}{n}\}$.  The energy and $z$ component of spin of the state $\ket{n+1,\nu}$ is $\varepsilon$ and $\hbar$, respectively,  higher than the state $\ket{n,\nu}$. According to \eq{eq:rho correlated spin- same temp}, the probability $P_{n,\nu}$ describing the occupation of the reservoir state $\ket{n,\nu}$ is proportional to $e^{-n\varepsilon/kT}$.  Taking into account the $\sbin{N}{n}$-fold degeneracy of this state then leads to the normalised probability distribution
\beq
   P_{n,\nu} = \frac{e^{-n\varepsilon/kT}}{(1+e^{-\varepsilon/kT})^N}
   \label{P_Boltzmann}
\eeq
which is independent of the index $\nu$.

As before the memory spin is in an unknown (maximally mixed) state and we wish to return it to the logical zero state. The first stage of our erasure scheme entails putting the memory spin in thermal and spin exchange contact with the reservoir, letting the combined system come to equilibrium, and then separating the memory spin from the reservoir. At this point the state of the memory spin is
\beq
    p_0\ket{0}\bra{0}+p_1\ket{1}\bra{1}
    \label{equil}
\eeq
with $p_1=e^{-\varepsilon/kT}/(1+e^{-\varepsilon/kT})=1-p_0$. We assume throughout that the reservoir is sufficiently large that the temperature $T$ is unaffected by the erasure process and that there are a large number of ancillary spins, all in the state$\ket{0}$, at our disposal. A CNOT operation \cite{Nielsen,Barnett} is then performed on the memory spin and one ancilla spin, with the former being the control qubit; this yields the state $p_0\ket{00}\bra{00}+p_1\ket{11}\bra{11}$ where $\ket{xy}$ represents the state $\ket{x}$ of the memory spin and $\ket{y}$ of the ancilla spin. The spin cost of this operation is $\hbar p_1= e^{-\varepsilon/kT}\hbar/(1+e^{-\varepsilon/kT})$ and the energy cost is $\varepsilon p_1=e^{-\varepsilon/kT}\varepsilon/(1+e^{-\varepsilon/kT})$. This system is again placed in thermal and spin exchange contact with the reservoir. The thermal and spin exchange between the reservoir and the memory-ancilla system is constructed to leave all states unchanged except for the following mapping
\beq
   \ket{2,1}\ket{00}\leftrightarrow\ket{0,1}\ket{11}
\eeq
where $\ket{n,\nu}\ket{ij}$ represents the reservoir collective state $\ket{n,\nu}$ and memory-ancilla system state $\ket{ij}$. The thermal and spin exchange continues for a sufficient time for the reservoir and memory-ancilla system to equilibrate. The state of the memory-ancilla system is then given by
\beq
    p_0\ket{00}\bra{00}+p_1\ket{11}\bra{11}
\eeq
where now $p_1=e^{-2\varepsilon/kT}/(1+e^{-2\varepsilon/kT})=1-p_0$. Another ancilla spin is added and a CNOT operation is performed as before to yield the state $p_0\ket{000}\bra{000}+p_1\ket{111}\bra{111}$ with spin and energy costs $e^{-2\varepsilon/kT}\hbar/(1+e^{-2\varepsilon/kT})$ and $e^{-2\varepsilon/kT}\varepsilon/(1+e^{-2\varepsilon/kT})$, respectively. The combined memory-ancilla system put in thermal and spin-exchange contact with the reservoir with the mapping
\beq
    \ket{3,1}\ket{000}\leftrightarrow\ket{0,1}\ket{111} \ .
\eeq

\begin{figure}  
\begin{center}
\includegraphics[width=70.4mm]{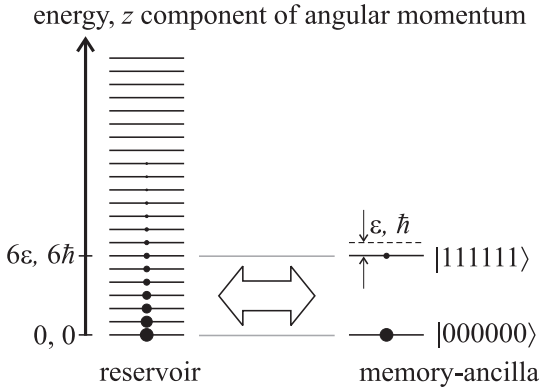}
\end{center}
\caption{Energy and angular momentum diagram for the erasure model with multiple costs.  The energy and the $z$ component of angular momentum of each system has discrete levels in multiples of $\varepsilon$ and $\hbar$, respectively.  For convenience, the lowest angular momentum level is labelled with zero angular momentum. To erase the contents of the memory, the reservoir and memory-ancilla systems are allowed to equilibrate by exchanging energy and angular momentum, next the gap between the states of the memory-ancilla system is increased, and then the process is repeated.  \label{fig2:energy+ang mom}
}

\end{figure}    

The process is repeated. Fig.~\ref{fig2:energy+ang mom} illustrates the situation after 5 cycles. After $m$ cycles, the memory-ancilla spins are in the logical zero state and the logical 1 state with probabilities $p_0$ and $p_1$ where $p_1=e^{-m\varepsilon/kT}/(1+e^{-m\varepsilon/kT})=1-p_0$. The upper limit for $m$ is the number of spins $N$ in the reservoir. In the limit of many repetitions and large $N$, the memory-ancilla system approaches a pure state where each spin is in the logical zero state $\ket{0}$.  In this limit, all the ancillary spins used in the process have been returned to their initial state $\ket{0}$ and the information represented by the initial state of the memory spin has been erased.  The total spin cost $\Delta J_z$ of the CNOT operations of the whole process approaches
\beq
   \Delta J_z=\sum_{n=1}^\infty
        \frac{e^{-n\varepsilon/kT}}{1+e^{-n\varepsilon/kT}}\hbar\ .
        \label{eq:cost J_z}
\eeq
This sum is bounded by
\beq
    \frac{kT\hbar}{\varepsilon}\ln(1+e^{-\varepsilon/kT}) < \Delta J_z <    \frac{kT\hbar}{\varepsilon}\ln(2)\ .
\eeq
If we include the spin of the initial state, then the spin cost is given by $\Delta J_z^\prime$ where
\beq
     \Delta J_z^\prime=\sum_{n=0}^\infty     \frac{e^{-n\varepsilon/kT}}{1+e^{-n\varepsilon/kT}}\hbar=\Delta J_z+\half\hbar\ ,
\eeq
\beq
    \frac{kT\hbar}{\varepsilon}\ln(2) < \Delta J_z^\prime < \frac{kT\hbar}{\varepsilon}\ln(1+e^{\varepsilon/kT})\ .
\eeq
Similarly the energy cost $\Delta E$ of erasing the memory of the memory spin is given by
\beq
    kT\ln(2) < \Delta E < kT\ln(1+e^{\varepsilon/kT})\ ,
\eeq
which includes the initial energy of the memory spin.

It is interesting to relate the two costs via dimensionless quantities.  Let $\alpha=e^{-\varepsilon/kT}/(1+e^{-\varepsilon/kT})$ be the probability that a reservoir spin is in the logical 1 state. The average energy of the reservoir spin is $\alpha\varepsilon$ and the average $\hat J_z$ value is $(\alpha-\half)\hbar$.  We can rewrite the lower bounds in terms of $\alpha$ by noting that
\beqa
    \frac{kT}{\varepsilon}&=&\frac{1}{\ln(\frac{1-\alpha}{\alpha})}
    \label{kT_epsilon}
\eeqa
so that
\beqa
    \Delta E &>& \frac{\varepsilon}{\ln(\frac{1-\alpha}{\alpha})}\ln(2)
    = kT \ln 2 \ ,
    \label{Delta_E}\\
    \Delta J_z^\prime &>& \frac{\hbar}{\ln(\frac{1-\alpha}{\alpha})}\ln(2)
     =\hbar \frac{kT}{\varepsilon}\ln 2\ .
     \label{Delta_J}
\eeqa
The first of these is the energy cost due to Landauer \cite{Landauer}.  In order to compare with our general result \eq{eq:general cost} we need to add appropriately scaled versions of these two costs as follows:
\beqa
    \beta\Delta E + \frac{\beta\varepsilon}{\hbar} \Delta J_z^\prime
     >2\ln 2   \ .
\eeqa
The right side is greater by a factor of two compared to that expected from \eq{eq:general cost} due to the fact that in order for the internal spin states of the reservoir to absorb the entropy of the memory, two different physical variables are changed and each is associated with a corresponding physical cost.  The actual change in the entropy of the reservoir is just $\ln 2$, however.  This example of information erasure is clearly not as efficient as it could be.

\section{Erasure without an energy cost} \label{sec:energy free}

\begin{figure}  
\begin{center}
\includegraphics[width=70.4mm]{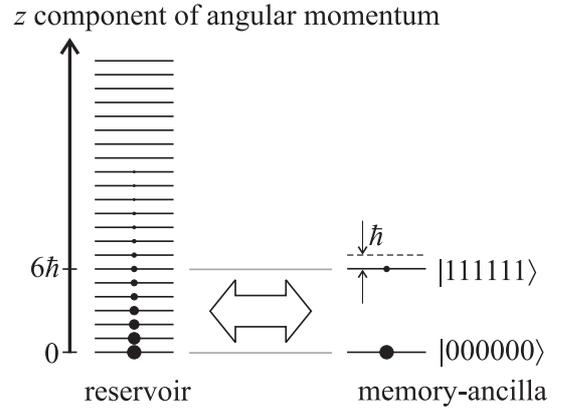}
\end{center}
\caption{Angular momentum diagram for the erasure model without an energy cost.  The $z$ component of angular momentum of the both systems have discrete levels in multiples of $\hbar$.  All states of both systems are degenerate in energy.  The reservoir and memory-ancilla systems are allowed to equilibrate by exchanging angular momentum only, next the gap between the states of the memory-ancilla system is increased, and then the process is repeated. \label{fig3:ang mom}
}
\end{figure}    

We have arrived at a strong link between the energy and angular momentum costs due to the relationship between the Zeeman splitting and the angular momentum.  Yet, it is not necessary for these energy and angular momentum to be so linked. By modifying, or better removing, the link between the the energy and angular momentum of the states it is possible to change the costs of erasure. In particular, if the magnetic field is removed, the logical states $\ket{0}$ and $\ket{1}$ become degenerate in energy and, as \eq{eq:H = J} indicates, the Hamiltonian term $\hat H_{in}$ vanishes. This means that only the expression on the right of \eq{eq:rho correlated spin} holds, i.e.
\beq
   \hat \rho =\exp[-(\mu+\beta \hat H_{ex}+\gamma \hat J_z)]
   \label{eq:rho spin only}
\eeq
where $\beta$ and $\gamma$ are independent Lagrange multipliers.   Examples of collections of spin particles that are described by density operators of this form are given by cold atomic gases confined in optical dipole force traps \cite{Pritchard}.  The internal states of the particles are described by the collective states $\ket{n,\nu}$ exactly as before, except that now they are associated with a probability $P_{n,\nu}$ that is proportional to $e^{-n\gamma\hbar}$ in accordance with to \eq{eq:rho spin only}. The normalised probability distribution is easily found to be
\beq
   P_{n,\nu}=\frac{e^{-n\gamma\hbar}}{(1+e^{-\gamma\hbar})^N}\ .
   \label{P_nm}
\eeq
We can relate the Lagrange multiplier $\gamma$ to the average of the $z$ component of spin of the reservoir by
\beq
  \gamma  = \frac{1}{\hbar}\ln(\frac{1-\alpha}{\alpha})\ .
  \label{gamma}
\eeq
where $\ip{\hat J_{z}}=(\alpha-\half)N\hbar$ for $0\le\alpha\le 1$ as before.  We also let the logical states of the memory and ancilla particle be energy degenerate.  The erasure process proceeds as in the previous section and is represented in Fig.~\ref{fig3:ang mom}. The only differences are that, due to the energy degeneracy of all states involved, the memory-ancilla system is brought into equilibrium with the reservoir by the exchange of spin angular moment only, the cost of performing the CNOT operations is also in terms of spin angular momentum and not energy, and the parametrisation of the probability distribution is in terms of $\gamma$ instead of $\beta$.  With this in mind we find that in the limit of many repetitions and large $N$, the memory-ancilla system approaches a pure state where each spin is in the logical zero state. The total angular momentum cost of the whole process is given by \eq{eq:cost J_z} with $\beta=1/kT$ replaced with $\gamma$, i.e. \cite{PRSA}
\beq
    \Delta J_z=\sum_{n=1}^\infty
    \frac{e^{-n\gamma\hbar}}{1+e^{-n\gamma\hbar}}\hbar
\eeq
which, correspondingly, is bounded by
\beq
   \gamma^{-1}\ln(1+e^{-\gamma\hbar}) < \Delta J_z <
   \gamma^{-1}\ln(2)\ .
\eeq
If we include the spin of the initial state, then the cost is
\beq
    \Delta J_z^\prime=\sum_{n=0}^\infty
    \frac{e^{-n\gamma\hbar}}{1+e^{-n\gamma\hbar}}=\Delta J_z+\half\hbar\ ,
\eeq
\beq
   \gamma^{-1}\ln(2) < \Delta J_z^\prime <
   \gamma^{-1}\ln(1+e^{\gamma\hbar})\ .
\eeq
Hence
\beq
   \Delta J_z^\prime > \gamma^{-1}\ln(2)
   \label{spincost}
\eeq
with the value of $\gamma$ given by \eq{gamma}. This is the total cost of the erasure; there is no energy cost in this case. {\it Clearly the costs associated with erasure depend on the physical nature of the memory system and the reservoir, and need not include an energy term, in contradistinction to the suggestion of Landauer and many others.}

This an important point and deserves some emphasis. Landauer's principle provides a basis for claiming an equivalence between thermodynamic and information entropies. But the equivalence rests on the erasure of information being {\em necessarily} associated with the dissipation of energy and with it the increase of a corresponding amount of thermodynamic entropy.  Our finding breaks this association.  As a consequence, information entropy can no longer be claimed to be equivalent to thermodynamic entropy.  Rather, how we might think of information entropy depends of the physical system used to store the information.  That we need to consider a physical system to store information is, however, consistent with Landauer's catchcry that `information is physical' \cite{InfoIsPhysical}.

We now examine the absence of the energy cost in more detail. Note that a cost of zero energy also occurs in Landauer's erasure when the reservoir temperature is zero. The erasure in this case can be thought of as a ``cooling'' process; the reservoir simply absorbs the heat of the memory bit. One may wonder if absence of an energy cost in the present case is associated with the reservoir being at zero temperature. However a degenerate reservoir in thermal equilibrium (at any temperature) has a flat probability distribution across all states and so cannot act to cool the memory bit.  We conclude it is not appropriate to consider the $T=0$ cooling mechanism for the degenerate case. We should rather think of the cost of erasure not in terms of energy but of the quantity defining the logic states.

Further insight into this issue is given by considering the non-degenerate model discussed in Section \ref{sec:multiple} in the degenerate limit $\varepsilon\to 0$ where the limit is approached in such a way that the $\Delta J_z$ cost given by \eq{Delta_J} is unchanged. The question we wish to address is whether the probability distribution in \eq{P_nm} can be consistently associated with a temperature $T=0$. To answer it, we examine the behaviour of $T$ as $\varepsilon\to 0$ such that the probability distributions in \eq{P_Boltzmann} and \eq{P_nm} are equal, i.e. \beq
   P_{n,\nu}=\frac{e^{-n\varepsilon/kT}}{(1+e^{-n\varepsilon/kT})^N}
   =\frac{e^{-n\gamma\hbar}}{(1+e^{-n\gamma\hbar})^N}\ ,
   \label{P_n_nu}
\eeq
and keeping the $\Delta J_z$ cost is fixed irrespective of the values of $T$ and $\varepsilon$. This means we need to keep the parameter $\gamma$ fixed, and thus from \eq{gamma} we need to keep $\alpha$ fixed. From \eq{kT_epsilon} we see that this requires $T\to 0$ linearly with $\varepsilon$.  In principle this could be achieved as follows. We could slowly reduce the Zeeman splitting $\varepsilon$ and at the same time reduce the temperature of the reservoir as $T=\varepsilon/k$. The angular momentum distribution of the reservoir, \eq{P_nm}, is unchanged in this process. One might then argue that \eq{P_nm} is therefore associated with $T=0$. We could, however, subsequently raise the temperature of the reservoir to some finite value $T>0$ while keeping the energy levels of the reservoir spins degenerate (i.e while maintaining a zero Zeeman splitting $\varepsilon=0$). As the spins are degenerate, no energy is exchanged between the external motional degrees of freedom and the internal spin degrees of freedom. Thus the probability distribution of the reservoir remains unchanged irrespective of the temperature of the reservoir. We could equally associate the distribution \eq{P_nm} with any arbitrary temperature.  This means that \eq{P_nm} cannot be consistently associated with any single temperature.  The $T=0$ cooling mechanism, therefore, is not operating here. We are thus led to the conclusion that energy (and thus work) does not play a role in erasure in the degenerate case.

The distribution \eq{P_nm}, moreover, represents a state of maximum entropy for a fixed $z$ component of angular momentum. Its ability to absorb the entropy of the memory spin is due to the maximization of entropy of the combined reservoir-memory-ancilla system subject to the conservation of $z$ component of angular momentum. Conservation of energy is trivially satisfied here due to the degeneracy and, as such, conservation of energy does not influence the equilibrium states. Again we are led to the conclusion that the $T=0$ cooling mechanism cannot operate in the degenerate case.

Finally, one may wonder whether the effect of a small residual magnetic field that induces a correspondingly small energy splitting $\varepsilon$ between the spin states would void our assertion that \eq{spincost} represents the total cost of the erasure. Indeed, \eq{Delta_E} indicates that an energy splitting of $\varepsilon$, however small, leads to an energy cost of $kT \ln 2$.  But any {\it known} magnetic field can be eliminated systematically, in principle.  We can assume, therefore, that any residual magnetic field is small and able to be estimated only. It will presumably vary from the site of one spin to another.  Consider the local residual field at the site of an arbitrary spin given by position vector ${\bf r}$.  Any component in the $x-y$ plane will induce precession of the spin orientation which will lead to inefficiencies and, thus, a higher spin cost or incomplete erasure.  It will not, however, fundamentally change the character of the cost of erasure.  In contrast, any component in the $z$ direction will produce a Zeeman splitting in energy, which we shall represent as $\varepsilon({\bf r})$ for a spin located at ${\bf r}$.  If the spins in the memory-ancilla system experience this splitting, the CNOT operation will incur an energy cost in addition to the angular momentum cost in \eq{spincost}.  The total energy cost of the CNOT operations will be given by
\beq
   \Delta E=\sum_{n=1}^\infty
        \frac{e^{-n\gamma\hbar}}{1+e^{-n\gamma\hbar}}\varepsilon({\bf r}_n)
\eeq
where ${\bf r}_n$  is the position of the ancilla spin which is the target of the $n$-th CNOT operation.  Let $\varepsilon({\bf r}_n)$ for any given $n$ be a stochastic variable with zero mean, i.e. $\overline{\varepsilon({\bf r}_n)}=0$ where the overline represents the ensemble average.  It then follows that
\beq
   \overline{\Delta E}=0\ .
\eeq
On average there is no net energy cost: the erasure process is as likely to result in an energy gain as an energy loss.  The fundamental point to be made here is that Landauer's cost of $kT\ln 2$ per bit is an average cost, and here the average energy cost is zero.  We conclude that a small, unpredictable, residual magnetic field does not void our result that, in principle, the cost of erasure is given in terms of angular momentum and not energy.

\section{Discussion}

We are now able to state {\em general} results.  To this end we write the costs of erasure in terms of a dimensionless parameter.  A convenient measure of the state of the reservoir for this is given by the number $n$ of particles that are in the logical $\ket{1}$ state. We can call this parameter the Hamming weight of the reservoir.  To keep our analysis general, we shall refer to the particles in the reservoir as qubits \cite{Barnett,Nielsen} rather than spins.  We have already seen that the increase in the entropy of the reservoir $\Delta S$ for the erasure of 1 bit is bounded below by $\ln(2)$, i.e.
\beq
   \Delta S > \ln(2)\ .
       \label{2nd_law}
\eeq
The actual change $\Delta S$ due to a Hamming cost of $\Delta n$ can be approximated to first order using
\beq
    \Delta S \approx \frac{dS}{dn}\Delta n\ .
\eeq
Combining these two results yields
\beq
    \Delta n > \ln(2)\left(\frac{dS}{dn}\right)^{-1}\ .
        \label{dSdn}
\eeq
For a reservoir comprising $N$ qubits with an average Hamming weight of $n=\alpha N$, the entropy is
\beq
    S=-N[(1-\alpha)\ln(1-\alpha)+\alpha\ln(\alpha)]
    \label{S}
\eeq
and so, on performing the derivative in \eq{dSdn}, we find the lower bound on the Hamming cost \beq
    \Delta n > \frac{\ln(2)}{\ln(\frac{1-\alpha}{\alpha})}
\eeq
which agrees with \eq{spincost} for the value of $\gamma$ given by \eq{gamma}.  Our erasure scheme is therefore optimal in this sense.

Similarly, if the logical states $\ket{0}$, $\ket{1}$ of the qubits are eigenstates of a physical variable $\hat V_k$ with corresponding eigenvalues $0$, $v_k$, then the average value of $\hat V_k$ for the reservoir is $\ip{\hat V_k}=\alpha v_k N$, and the entropic cost can be reexpressed as
\beq
    \Delta S \approx \frac{dS}{d\ip{\hat V_k}}\Delta \ip{\hat V_k} \ .
\eeq
Using \eq{2nd_law} and \eq{S}, we then find the corresponding cost for arbitrary erasure schemes
\beq
    \Delta \ip{\hat V_k} > \frac{v_k\ln(2)}{\ln(\frac{1-\alpha}{\alpha})} \ .
\eeq
The total cost of erasure is therefore
\beq
    \ln\left(\frac{1-\alpha}{\alpha}\right)\sum_{k=1}^M\frac{1}{v_k}\Delta V_k > M\ln(2)
\eeq
for $M$ physical variables. The right side is $M$ times the right side of \eq{eq:general cost} due to the fact that here the expectation values $\ip{\hat V_k}$ vary in proportion to $\alpha$, whereas in \eq{eq:general cost} the expectation values $\ip{\hat V_k}$ are assumed to vary independently.  Information erasure is clearly inefficient when more than one physical variable is associated with the degree of freedom of the reservoir that absorbs the entropy of the memory.

These results open up a range of topics for investigation.  For example, the operation of Carnot `heat' engines operating with angular momentum reservoirs and generating angular momentum `work' (or some other resource) instead of mechanical work.  Another possibility is the use of a {\em combination of different types of reservoir}. For example, a Maxwell's demon can operate on a single thermal reservoir to extract work from the reservoir. However there is an associated unmitigated cost in that the memory of the demon has to be erased. Bennett's argument \cite{bennett1982} is to use a thermal reservoir following Landauer's erasure principle to do this, so that the extracted work is (more than) balanced by the cost of erasure. However, given the forgoing, we now know that the memory of the demon can be erased using an entirely different reservoir at no cost in energy.  The cost instead could be in terms of angular momentum, say \cite{PRSA}.  Fundamentally, however, the irreducible cost is information theoretic, i.e. a cost in terms of Shannon entropy. This suggests the operation of generalized Carnot cycles between different kinds of reservoirs. The fundamental principle of operation being a movement of entropy from one reservoir to another.  These issues will be explored elsewhere.

\acknowledgements{SMB thanks the Royal Society and the Wolfson Foundation for financial support.  JAV thanks T. Rudolph, D. Jennings, S. Jevtic and M. Lostaglio for useful discussions, and gratefully acknowledges financial support from the ARC Centre of Excellence Grant No. CE110001027.
}


%


\end{document}